\documentstyle[epsf,preprint,aps,eqsecnum,amsfonts]{revtex}
\begin{document}

\draft

{\tighten
\preprint{\vbox{\hbox{WIS-94/39/Sep-PH}
                \hbox{CALT-68-1952}
                \hbox{hep-ph/9409418} }}

\title{Transverse tau polarization in inclusive
       $\bar B\to\tau\,\bar\nu\,X$ decays}

\author{Yuval Grossman$^a$ and Zoltan Ligeti$^{a,b}$\thanks{%
Present address.} }

\address{ \vbox{\vskip 0.5truecm}
  $^a$Department of Particle Physics \\
      Weizmann Institute of Science, Rehovot 76100, Israel \\
\vbox{\vskip 0.1truecm}
  $^b$California Institute of Technology, Pasadena, CA 91125 }

\maketitle

\begin{abstract}%
We calculate, in the framework of multi Higgs doublet models, the $CP$
violating transverse tau polarization, $P_\perp$, in inclusive semileptonic $B$
meson decays.  We find that $P_\perp$ diverges at ${\cal O}(1/m_b^2)$ of the
heavy quark expansion.  We discuss the physical reasons for this divergence and
show how to regularize it.  We find large $1/m_b^2$ corrections that can
suppress the $CP$ asymmetry by as much as 30\%.  The maximal allowed
asymmetry is $a_{CP}\sim0.34$.  We discuss how the allowed range is expected to
change in the future.
\end{abstract}

}%end tighten

\newpage
\narrowtext

\def\lo{\lambda_1}
\def\lt{\lambda_2}
\def\src{\hat m_c}
\def\srt{\hat m_\tau}
\def\hqs{{\hat q}^2}
\def\rr{r_{\rm Re}}
\def\ri{r_{\rm Im}}

\section{Introduction}

Multi Higgs doublet models (MHDM) allow for new $CP$ violating phases in
charged scalar exchange\cite{Lee}.  In models with natural flavor conservation
(NFC)\cite{GlWe} at least three Higgs doublets are needed for the scalar sector
to allow $CP$ violation \cite{Wein}.  The effect cannot be large in $CP$
asymmetries in neutral $B$ decays or in $CP$ violation in the $K-\bar K$ system
\cite{gn}.  However, there could be significant contributions to processes for
which the SM prediction is small: the electric dipole moment of the neutron
\cite{EDM}, $CP$ asymmetries in top decays\cite{topdecay}, and $CP$ violation
in semileptonic decays.

As triple-vector correlation is odd under time-reversal, the experimental
observation of such correlation would signal $T$ and -- assuming $CPT$ symmetry
-- $CP$ violation.  In three body semileptonic decays the lepton transverse
polarization is such an observable.  It cannot be generated by vector or
axial-vector interactions only \cite{Miriam}, so it is particularly suited for
searching for $CP$ violating MHDM contributions.  The muon transverse
polarization has been studied in $K\to \pi\,\mu\,\nu$ decays
\cite{Miriam,allk}.  Since in models with NFC scalars couple more strongly to
heavier fermions, the expected signals are stronger in heavy quark decays
\cite{allheavy,GolVal,Eilam,Garisto}.

For top decays the quark level description is valid, as the top quark is
expected to decay before it hadronizes.  This is, however, not the case for $B$
decays.  Recently, it has been observed that inclusive semileptonic decays of
hadrons containing a single heavy quark allow for a systematic, QCD-based
expansion in powers of $1/m_Q$ \cite{CGG}.  The heavy quark limit
coincides with the free quark decay model and there are no
corrections to this result at order $1/m_Q$ \cite{CGG,BUV}.  The leading
nonperturbative corrections are of order $1/m_Q^2$ and depend on only two
hadronic parameters, which parameterize forward matrix elements of local
dimension-five operators in the heavy quark effective theory (HQET).
These corrections have been computed for a number of processes
\cite{BUV,all,us,else,YZ}.

Using the combination of the operator product expansion and HQET, semileptonic
$B$ meson decays into a tau lepton have been studied in the framework of the SM
\cite{us,else} and MHDM \cite{YZ}.  In this paper we use this method to study
the tau transverse polarization in the framework of MHDM.  In addition to
finding the correction to the spectator model result, this calculation has some
interesting technical aspects.  The nonanaliticity of the triple product
correlation on the boundaries of the Dalitz plot gives rise to infinities in
the
calculation.  Such divergencies do not appear in the calculations of either
cross-sections or longitudinal polarization.  We discuss their physical meaning
and how to eliminate them.

\section{The calculation}

We consider a general multi Higgs doublet model with NFC (for a recent analysis
and notation see \cite{Yuval}).  In order to have observable $CP$ violation,
the lightest charged scalar has to be much lighter than the heaviest one
\cite{Lav}.  Here we assume that the all but the lightest charged scalars
effectively decouple from the fermions.  The Yukawa couplings of the lightest
charged scalar to up-type quarks, down-type quarks and charged leptons are
determined by the parameters $X$, $Y$ and $Z$, respectively.  The terms in the
effective Lagrangian relevant for $\bar B\to\tau\,\bar\nu\,X$ decays are
\begin{equation}\label{lag}
{\cal L} = -V_{cb}\,{4G_F\over\sqrt2}\,
  \Big[ (\bar c\,\gamma^\mu P_L\, b)\, (\bar\tau\,\gamma_\mu P_L\,\nu_\tau)
  - R\,(\bar c\, P_R\, b)\, (\bar\tau\, P_L\,\nu_\tau) \Big] \,,
\end{equation}
where
\begin{equation}\label{Rdef}
R = r^2\, m_\tau\, m_b^Y \,, \qquad r^2 = XZ^*/m_H^2\,,
\end{equation}
$P_{R,L}=\frac12(1\pm\gamma_5)$, and $m_H$ denotes the mass of the lightest
charged scalar particle.
The running quark masses are denoted by an upper index $Y$.
The first term in Eq.~(\ref{lag}) gives the SM contribution, while
the second one is that of the charged scalar.  We neglect a term proportional
to $Y\,m_c^Y$: first, it is suppressed by the mass ratio $m_c^Y/m_b^Y$; second,
$Y$ is bounded to be of ${\cal O}(1)$, while $X$ can be large\cite{Yuval}.

The tau transverse polarization, $P_\perp$, in the decay
$\bar B\to\tau\,\bar\nu\,X$ is defined as the tau polarization
component along the normal vector of the decay plane.  It is given by
\begin{equation}
P_\perp = \frac{\vec s_{\tau} \cdot (\vec p_{\tau} \times \vec p_X)}
         {|\vec p_{\tau} \times \vec p_X|}\,,
\end{equation}
where $\vec s_{\tau}$ is the tau spin three vector and $\vec p_{\tau}$
($\vec p_X$) is the three-momentum of the tau lepton (hadron).
We define $\Gamma^+$ ($\Gamma^-$) as the rate of finding the tau spin
parallel (anti-parallel) to the normal vector of the decay plane.
The total rate $\Gamma=\Gamma^+ +\Gamma^-$ has been calculated in \cite{YZ}.
We then have to calculate only $\Gamma^+ - \Gamma^-$.

The techniques of the calculation are described in detail elsewhere \cite{all},
so we give only the necessary details, followed by the results of the
computation.  The asymmetry arises only from the interference between the
$W$ mediated and the Higgs mediated diagrams.  The interference term in the
hadronic current can be decomposed as
\begin{equation}
(2\pi)^3 \sum_X \delta^4(p_B-q-p_X)\,
  \langle B_v|\,J^\dagger\, |X\rangle\, \langle X|\,J^\mu\, |B_v\rangle\,
  = v^\mu\,U_1 + q^\mu\,U_2 \,,
\end{equation}
where
\begin{equation}
J^\mu = \bar c\, \gamma^\mu P_L\, b\,, \qquad
J = {1\over m_H}\, \bar c\, m_b^Y X P_R\, b\,,
\end{equation}
$q^\mu$ is the momentum of the lepton pair, and $v^\mu$ is the four-velocity of
the decaying $B$ meson.  We work in the $B$ rest frame, where $\theta$ is the
angle between $\vec p_{\tau}$ and $\vec p_X$, and the tau spin projection is
along the four-vector $s=(0,0,{\rm sgn}[\sin\theta],0)$.  Upon contraction of
the lepton current with $W^\mu$, the term proportional to $U_2$ vanishes and
the $U_1$ part yields the differential asymmetry
\begin{equation} \label{difacp}
{{\rm d}(\Gamma^+-\Gamma^-)\over{\rm d}\hqs\,{\rm d}x\,{\rm d}y} =
  {|V_{cb}|^2\, G_F^2\, m_b^5 \over 16\pi^3}\, U_1\,{\rm Im}R\,
  \sqrt{xy(\hqs-\srt^2)-x^2\srt^2-(\hqs-\srt^2)^2}\,.
\end{equation}
where we defined dimensionless variables
\begin{equation}
y = {2E_\tau\over m_b}\,,\qquad
x = {2E_\nu\over m_b}\,,\qquad
\hqs = {q^2\over m_b^2}\,,\qquad
\src = {m_c\over m_b}\,,\qquad
\srt = {m_\tau\over m_b}\,,
\end{equation}
and $E_\tau$ ($E_\nu$) is the energy of the tau (neutrino).
The OPE yields for $U_1$ \cite{YZ}
\begin{equation}
U_1 = \delta(\hat z) + {1\over6m_b^2} \left\{
  [5(\lo+3\lt)(x+y) -36\lt]\, \delta^\prime(\hat z) +
  \lo[4\hqs-(x+y)^2]\, \delta^{\prime\prime}(\hat z) \right\}\,,
\end{equation}
where $\lo$ and $\lt$ are parameters of the HQET, $\hat z=1+\hqs-\src^2-y-x$,
and the prime means ${\rm d}/{\rm d}\hat z$.

In order to obtain the total asymmetry we have to integrate (\ref{difacp}) over
$x$, $y$, and $\hqs$.  This integral diverges because the square root function
is not analytic.  The derivatives resulting from the $\delta^\prime$ and
$\delta^{\prime\prime}$ terms yield negative powers of the square root in
(\ref{difacp}) to be evaluated precisely at their pole.  Physically, the reason
for this non-analyticity is that the orientation of the decay plane changes
when the tau and the hadron momenta become collinear.  At these points in
phase-space (corresponding to the boundary of the Dalitz plot in the $y-\hqs$
plane) the decay plane is not well-defined, so neither is the transverse
polarization.  In order to regularize this divergence, we have to rely on
the physical picture.
Experimentally the region near the boundary of the Dalitz plot is not
accessible: any finite resolution gives a region in which the transverse tau
polarization is not measurable as even the decay plane cannot be reconstructed.
This suggests to regularize the divergencies by a simple cutoff, and then take
the limit of the cutoff approaching the border of the Dalitz plot.
Another possibility is to replace the square root by an analytic function,
which means physically to make the flip of the orientation of the decay plane
smooth.  It is convenient to apply the following replacement
\begin{equation}\label{reg}
\sqrt{f(x,y,\hqs)} \to\sqrt{f(x,y,\hqs)}\,
\left[1-e^{-nf^2(x,y,\hqs)}\right]\,,
\end{equation}
and interchange the order of the integration with the $n\to\infty$ limit.
This procedure eliminates the infinities.  In fact, this prescription is only
sensitive to the derivatives of the regularized function at the points where
the original square root function vanished (these points constitute the
border of the Dalitz plot).  Since at these points all derivatives of
(\ref{reg}) vanish, this should reproduce the cutoff result.  We checked
that in the $m_c\to0$ limit these two procedures indeed give identical results.
(The calculation using a cutoff becomes unmanageable when we keep $m_c$
finite.)  Technically the replacement (\ref{reg}) is much simpler than
doing the full calculation with a cutoff.

After the integration we obtain for the $CP$ violating asymmetry
\begin{equation}
a_{CP} = {\Gamma^+ - \Gamma^- \over \Gamma^+ + \Gamma^-} = {1\over\Gamma}\,
  {|V_{cb}|^2\, G_F^2\, m_b^5 \over 192\pi^3}\, {\rm Im}R\,
  \tilde\Gamma_\perp\,,
\end{equation}
where
\begin{eqnarray}
\tilde\Gamma_\perp = \frac{8\pi}{35}\, \bigg\{ C^4 &&
  \Big[1+4\src-4\src^2-\src^3 + 4\srt(1+5\src+\src^2) - 4\srt^2(1-\src)
  - \srt^3 \Big] \nonumber \\
+ {\lo\over 6m_b^2}\, C^3 &&
  \Big[3(1-\src)^2(8+5\src+\src^2) + 9\srt(8-3\src-4\src^2-\src^3) \nonumber\\
  && +4\srt^2(8-9\src-6\src^2) + 9\srt^3(1-\src) + 3\srt^4 \Big] \nonumber \\
+ {\lt\over 2m_b^2}\, C^3 &&
  \Big[3(-16-13\src+9\src^2+15\src^3+5\src^4)
  - 9\srt(16+29\src+20\src^2+5\src^3) \nonumber\\
  && -4\srt^2(2+45\src+30\src^2) + 45\srt^3(1-\src) + 15\srt^4 \Big] \bigg\}\,,
\end{eqnarray}
and $C=(1-\src-\srt)$.
In the limit $\lambda_1,\lambda_2\to0$, corresponding to the free quark
decay model, our results agree with Ref.\cite{Eilam}.
(Although Ref.\cite{Eilam} works in the tau rest frame, our definitions
of the asymmetry are identical.)

\section{Discussion}

In our numerical analysis we use the same input parameters as in Ref.\cite{YZ}.
Let us further define:
\begin{equation}
\rr = \sqrt{|{\rm Re}\,(r^2)|} = \frac{\sqrt{|{\rm Re}\,( XZ^*)|}}{m_H}
\,, \qquad
\ri = \sqrt{|{\rm Im}\,(r^2)|} = \frac{\sqrt{|{\rm Im}\,( XZ^*)|}}{m_H}\,.
\end{equation}
The quantity $\ri$ enters the asymmetry $a_{CP}$ through its numerator
quadratically, and the denominator in a more complicated manner.
The dependence of
$a_{CP}$ on $\rr$ is only through the total decay rate.  In our analysis we use
only the allowed values of $\ri$ and $\rr$, and we discuss the present
constraints on them below.  By varying these parameters within their allowed
ranges, we find the allowed values of $a_{CP}$ as a function of $\ri$. Our
result is given by the shaded region between the solid lines in Fig.~1.

A few points are in order regarding this result:

$a$.  The $1/m_b^2$ corrections to the free quark decay model decrease $a_{CP}$
for all values of $\ri$.  The numerator is suppressed by about 25\%.  For small
values or $\ri$ the denominator is also suppressed and the total correction is
small.  However, for large values of $\ri$ the total branching ratio is
enhanced \cite{YZ} and $a_{CP}$ can be suppressed by as much as 35\%.

$b$.  The uncertainty in our result (corresponding to the width of the shaded
region in Fig.~1) takes into account the theoretical uncertainties as in
Ref.\cite{YZ}.

$c$.  The improvement of our calculation over the spectator model originates
from large $1/m_b^2$ corrections, and from using $m_b$ and $m_c$ as determined
by the HQET rather than treating them as independent input parameters.  To
illustrate this, we plotted with dashed lines in Fig.~1 the prediction of the
free quark decay model corresponding to $1.4<m_c<1.5\,{\rm GeV}$ and
$4.6<m_b<5\,{\rm GeV}$.

$d$. Even in the SM ``fake" asymmetries arise as a background due to $CP$
conserving unitary phases from final state interaction (FSI) and from real
intermediate states.  These are expected to be less than 1\% \cite{KSW}.
Furthermore, they can be removed using the fact that asymmetries due to $CP$
violating phases have different sign in $B$ and $\bar B$ decays, while unitary
phases yield asymmetries with the same sign \cite{OkKh}.  Therefore, in our
analysis we ignore them.

The strongest bound on $\ri$ comes at present from the measurement of the
(unpolarized) $\bar B\to \tau\,\bar\nu\,X$
branching ratio.  The total decay rate $\Gamma$ has the form
\begin{equation}\label{rate}
\Gamma = W - I\,{\rm Re}R + H\,|R|^2
\end{equation}
where $W$ ($I$, $H$) denotes the SM (interference, Higgs) term.  This is a
circle in the complex $\{\rr,\ri\}$ plane, centered on the $\rr$ axis.  In our
analysis we set the value of $\rr$ to allow for maximal $\ri$ subject to the
experimental constraint.  From the measurements \cite{ALEPH,L3,PDG}
\begin{eqnarray}
{\rm BR}\,(\bar B\to \tau\,\bar\nu\,X) &=& 2.98\pm0.47\,\%\,, \nonumber\\
{\rm BR}\,(\bar B\to \ell\,\bar\nu\,X) &=& 10.43\pm0.24\,\%\,,
\end{eqnarray}
and using Ref.\cite{YZ} we obtain at the $2\sigma$ level
\begin{equation} \label{imrbound1}
\ri < 0.47 \,{\rm GeV}^{-1}\,.
\end{equation}

Due to the negative sign of the interference term in (\ref{rate}), the
strongest possible bound from the total branching ratio that can be obtained
in the future is about $\ri\lesssim0.35 \,{\rm GeV}^{-1}$.
The longitudinal polarization of the tau in  the same decay has a better
sensitivity \cite{YZ,Kali} down to about $\ri\lesssim0.25 \,{\rm GeV}^{-1}$.
Note, that from
just one $CP$ conserving observable no determination of $\ri$ is possible,
since deviation from the SM prediction can also be induced by $\rr$.  In
principle, however, by comparing two observables with different dependence on
$\rr$ and $\ri$ one may be able to detect a nonzero $\ri$.

The purely leptonic decay $\bar B\to \tau\,\bar\nu$ is also sensitive to $\ri$.
Using Ref.\cite{Hou} and the recent ALEPH result \cite{ALEPH}
\begin{equation}
{\rm BR}\,(\bar B\to \tau\,\bar\nu) < 1.5 \times 10^{-3} \qquad
  ({\rm 90\%\, CL}) \,,
\end{equation}
together with the $f_B>140\,{\rm MeV}$ and $|V_{ub}|>0.0024$ bounds, we get
\begin{equation}
\ri < 0.58  \,{\rm GeV}^{-1}\,.
\end{equation}
Due to the dependence on $\ri$ and the large theoretical uncertainties in the
calculation we estimate that this bound cannot become stronger in the future
than $\ri\lesssim0.3\,{\rm GeV}^{-1}$.

The bound on $\ri$ from $P_\perp$ of the muon in $K\to \pi\,\mu\,\nu$ decays
is obtained by converting the measured quantity, ${\rm Im}\xi$, to $\ri$
\cite{Kuno}
\begin{equation} \label{imbound}
|{\rm Im}\,\xi| \approx \ri^2\, m_K^2 \,.
\end{equation}
The theoretical uncertainty in this relation enters through the calculation of
certain form factors.  ${\rm Im}\,\xi$ has been measured in both charged and
neutral kaon decays.  The background of these measurements due to $CP$
conserving FSI phases are different.  For the charged kaon decay
$K^+\to \pi^0 \,\mu^+ \,\nu$, FSI are second order in the
electromagnetic interaction and therefore very small: ${\cal O}(10^{-6})$
\cite{kpfsi}.  In neutral kaon decay $K_L\to \pi^- \,\mu^+ \,\nu$, there are
two charged particles in the final state, and so FSI is first order in the
electromagnetic interaction and can contribute to ${\rm Im}\,\xi$ at the
${\cal O}(10^{-3})$ level.  Thus, in order to set a bound on $\ri$ we use
${\rm Im}\,\xi(K_L\to \pi^- \,\mu^+ \,\nu) \approx
{\rm Im}\,\xi^{FSI} + \ri^2\, m_K^2$ instead of Eq.~(\ref{imbound}),
where ${\rm Im}\,\xi^{FSI} \approx 0.008$ \cite{OkKh,Kfsi}.
The PDG data \cite{PDG}
\begin{eqnarray}
{\rm Im}\,\xi(K_L\to \pi^- \,\mu^+ \,\nu) &=& -0.007\pm0.026 \,, \nonumber\\
{\rm Im}\,\xi(K^+\to \pi^0 \,\mu^+ \,\nu) &=& -0.017\pm0.025 \,,
\end{eqnarray}
yields at the $2\sigma$ level
\begin{equation} \label{imrbound2}
\ri < 0.47  \,{\rm GeV}^{-1}\,.
\end{equation}
Numerically, the bounds (\ref{imrbound1}) and (\ref{imrbound2}) are the same.
However, we consider the bound (\ref{imrbound2}) less reliable as it is
sensitive to the uncertainties in the form factors and in the FSI phases in
$K_L$ decay.

In the near future the E246 experiment at KEK will measure the transverse muon
polarization in $K^+\to \pi^0\,\mu^+\,\nu$ decay.  They are aiming at a
sensitivity to $P_\perp$ of about $5\times 10^{-4}$ which corresponds to
$\ri\approx 0.08\,{\rm GeV}^{-1}$~\cite{Kuno}.  From Fig.~1 we see that in
order to achieve a better sensitivity to $\ri$ from the measurement of
$P_\perp$ of the tau in $\bar B\to \tau\,\bar\nu\,X$ decay, one needs to
measure $P_\perp$ at an accuracy of better than $2\%$.  It is important whether
$P_\perp$ of the tau can be reconstructed at all from its decay products in $B$
decays, either at LEP, at $B$ factories, or in hadron colliders.  To our
knowledge, this issue is not yet settled and further experimental studies are
called for.

One may naively expect that for the same $CP$ violating parameters, the tau
polarization in $B$ decays would be $m_b\,m_\tau/m_s\,m_\mu\sim 500$ times
larger than the muon polarization in $K$ decays.  However, our calculation
gives only about a factor of 40 enhancement.  The scalar matrix element in
exclusive $K$ decays is enhanced by a factor of $(m_K/m_s)^2 \sim 7.6$.  The
finite mass effects in the phase space integrals reduce $a_{CP}$ by about 45\%
in $B$ decays, while by only 30\% in $K$ decays.  Finally, the $1/m_b^2$
corrections reduce $a_{CP}$ in $B$ decays by up to 35\%.  In addition, the
experimental measurement of the transverse muon polarization in $K$ decays is
simpler than measuring the tau polarization in $B$ decays.  Thus we conclude
that although most of the MHDM parameters are easier to probe in processes
involving heavy fermions, it may be easier to constrain $\ri$ in kaon decays.

It is interesting to compare our results to Ref.\cite{Garisto}, which considers
exclusive $B$ decays.  There it was found that the allowed ranges for $a_{CP}$
extend from ${\cal O}(35\%)$ in $B\to D^*\,\tau\,\bar\nu$ decay up to
${\cal O}(1)$ in the $B\to D\,\tau\,\bar\nu$ channel.  We believe that the
suppression that we found in the inclusive channel is likely to apply to the
exclusive ones as well.  On the one hand, it would be easier to measure
$a_{CP}$ in inclusive decays (larger statistics, no need to reconstruct
specific hadronic final states) than in exclusive channels, and the exclusive
calculation also suffers from larger theoretical uncertainties.  On the other
hand, the results of Ref.\cite{Garisto} suggest that the asymmetry in the
$B\to D\,\tau\,\bar\nu$ decay may be enhanced.

Finally, our analysis holds with minor modifications for inclusive
$\bar B\to\mu\,\bar\nu\,X$ and $D\to \mu\,\bar\nu\,X$ decays as well.
However, the
possible signals are smaller and harder to measure experimentally than in the
$\bar B\to\tau\,\bar\nu\,X$ and $K\to \pi\,\mu^+\,\nu$
decay modes discussed above.

\section{Summary}

We studied the transverse tau polarization asymmetry in inclusive
$\bar B\to\tau\,\bar\nu\,X$ decays in MHDM.  We use the heavy quark expansion
to incorporate the nonperturbative $1/m_b^2$ corrections.  Unlike previous
cases, in the present calculation divergencies appear that have to be carefully
removed.  The $1/m_b^2$ corrections turn out to be important and reduce the
asymmetry by as much as 30\% in certain regions of the parameter space.  We
studied the present and possible future experimental bounds on the $CP$
violating parameter of the Higgs sector, $|{\rm Im}(XZ^*)|/ m_H^2$.  The
strongest bound comes at present from the measured $\bar B\to\tau\,\bar\nu\,X$
branching ratio.  In the future -- even if a new source of $CP$ violation will
not be discovered by E246 at KEK -- the bound on the measurement of the muon
transverse polarization in $K\to \pi\,\mu\,\nu$ will provide the strongest
constraint.

\acknowledgements
We thank Yossi Nir and Mark Wise for discussions and comments on the
manuscript, Yael Shadmi and Neil Marcus for helpful conversations.
Z.L.\ was supported in part by the U.S.\ Dept.\ of Energy under Grant no.\
DE-FG03-92-ER~40701.

{\tighten

\begin{figure}
\epsfysize=15truecm
\centerline{\epsfbox{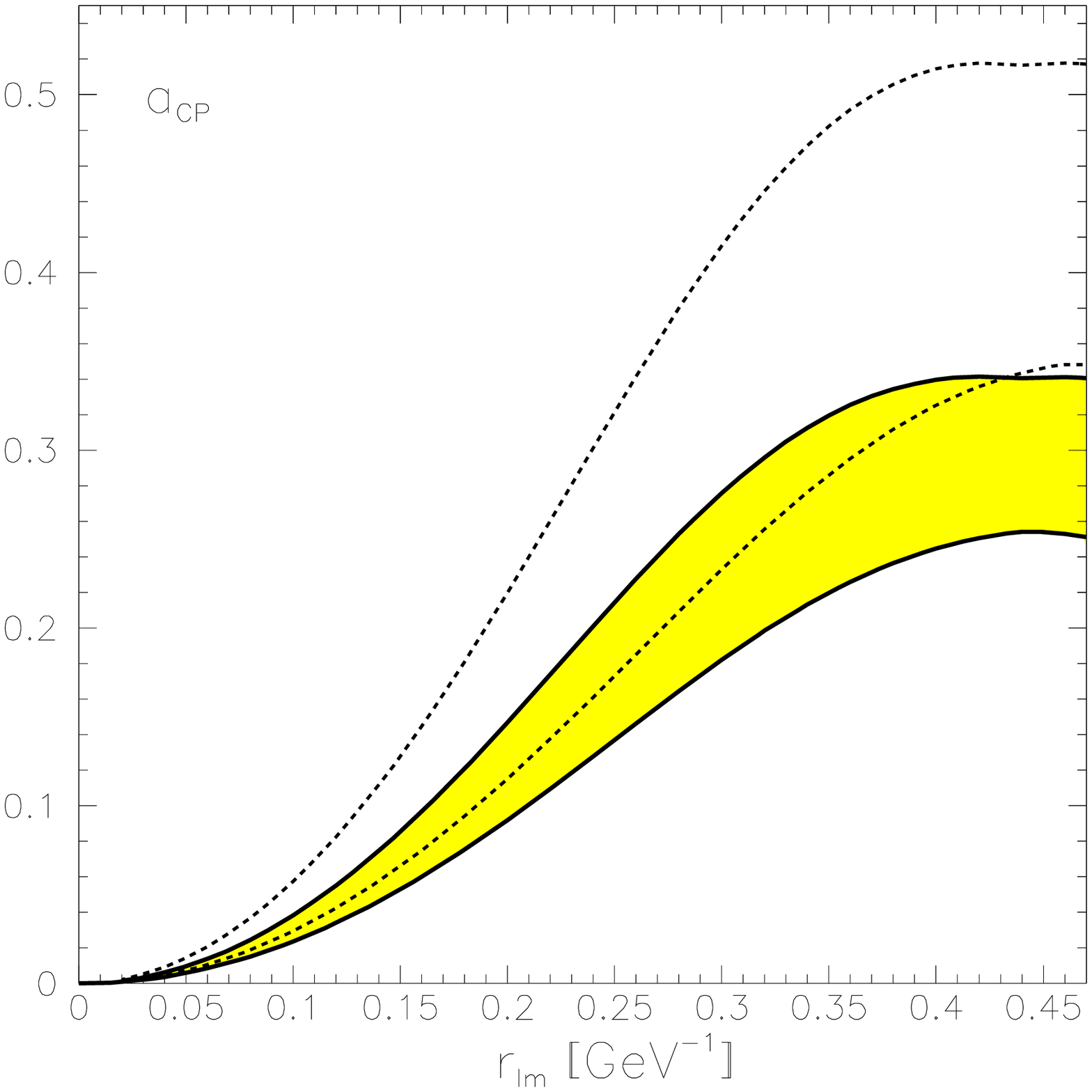}}
\caption[a]{
$a_{CP}$ as a function of $\ri=
\sqrt{|{\rm Im}\,(r^2)|}= \sqrt{|{\rm Im}\,(X Z^*)|}/{m_H}$.
The shaded area between the
solid lines is our result.  The area between the dashed lines gives the
free quark decay model result. }
\end{figure}

} %end tighten (references & figure captions)

\end{document}